\documentclass[%
 reprint,
 amsmath,amssymb,
 aps,
 pre,
]{revtex4-2}

\usepackage{graphicx}
\usepackage{dcolumn}
\usepackage{bm}

\begin{document}


\title{Contact angle hysteresis and static friction for two-dimensional droplets}

\author{Jong-In Yang}
 \email{jiyang@hanyang.ac.kr}
\affiliation{%
Department of Applied Physics, Hanyang University ERICA, 55 Hanyangdeahak-ro, Sangnok-gu, Ansan, Gyeonggi-do, 15588, Republic of Korea
}%
\author{Jooyoo Hong}%
 \email{jhong@hanyang.ac.kr}
\affiliation{%
Department of Applied Physics, Hanyang University ERICA, 55 Hanyangdeahak-ro, Sangnok-gu, Ansan, Gyeonggi-do, 15588, Republic of Korea
}%




\date{\today}

\begin{abstract}
Contact angle hysteresis of droplets will be examined in light of static friction between liquid drop and solid surface. Unlike frictions in solid-solid interfaces, pinning forces at contact points or contact lines would be the cause of friction. We will define coefficients of static friction and relate them with advancing and receding contact angles for the case of two-dimensional droplets. In our work sessile drops in an inclined plane, and pendent drops in a slanted ceiling will all be analyzed within a single framework with the inclination angle as a free parameter. We can then visualize the gradual change of shapes of a droplet put on a plane as the inclination angle changes adiabatically to make a complete turn. We point out that there could be two distinct stable configurations of pendent droplets for the same physical conditions, hence dubbed the bifurcation. And in the case of pendent droplets we investigate at what ranges of parameters pinch-offs occur, and find an interesting relation between the fallen-out volume and the Bond number.
\end{abstract}

\maketitle


\section{Introduction}
While a liquid drop lying on a solid surface is an unremarkable object we come across in everyday life, it has brought about an extensive list of research for over two centuries. And we still do not have a complete understanding of all its fascinating aspects. At a contact line, more precisely at a cross-sectional point on a contact line where gas, liquid, and solid come together, the so-called Young's equation \cite{Thomas_Young_1805, Lord_Kelvin_1886, Lamb_1928, Berry_1971, Berry_1975, Finn_1986, Myshkis_1987, Safran_1994, Pozrikidis_2016, Landau_1987, Langbein_2002, Berthier_2012, Shkolnikov_2019, Joanny_1984, deGennes_2003, Drelich_2020, Butt_2022, Daniel_2023} has been known to hold
\begin{equation*}
    \gamma^{sg} - \gamma^{sl} = \gamma^{lg} \cos \theta_{Y} \ .
\end{equation*}
Here $\gamma^{sg}$, $\gamma^{sl}$, and $\gamma^{lg}$ are the interfacial tension between solid-gas, solid-liquid, and liquid-gas, respectively. And $\theta_{Y}$ is the Young's angle. This relation comes as a consequence of force balance in the direction tangential to the solid surface when there is no external force other than surface tensions. However, when a liquid drop at rest is acted by an additional external force, the contact line does not move immediately. It would not budge until the force exceeds a certain critical value, which means the modification of the force balance and thus the change of Young's angle. This phenomenon is well known and dubbed contact angle hysteresis. Though there are no shortage of works on it in the literature \cite{Lv_2018, Tenan_1982, Nosonovsky_2015, Good_1952, Santos_2011, Makkonen_2017, Heijden_2018, Adam_Jessop_1925, Extrand_Gent_1990, Carre_1995, Odunsi_2023, Law_2022, YijieXiang_2022, McHale_2022, Lindeman_2023, Zhang_2023, Macdougall_1942, Yarnold_1949, Furmidge_1962, Shanahan_1982, Shanahan_1984, Joanny_1984, Langbein_2002, deGennes_2003, Krasovitski_2005, Kulinich_2009, Jung-yeulJung_2010, HyunsooSong_2011, Bormashenko_2011, Belman_2012, Berthier_2012, Ferrari_2012, Bormashenko_2013, Eral_2013, Pittoni_2014, White_2015, Shkolnikov_2019, Drelich_2020, Tadmor_2021, Butt_2022, Daniel_2023} there lacks a unified and systematic approach to the understanding of this interesting problem. 

Since the contact angle determines the shape of a droplet together with the surface tensions and the volume (or the cross-sectional area) as well as the topography of the substrate, it is important to understand and determine the contact angle in as precise a way as possible. A liquid drop in reality is an object in space, but to understand it we have to solve the governing equation numerically since it does not allow analytical solutions, which hampers a systematic understanding. In contrast, a 2-dimensional droplet or a long droplet with translational symmetry along an axis, though idealized and approximate, allows one to deal with this age-old problem as thoroughly as possible, which is the subject of the present work.

There exists a similar work by Lv and Shi \cite{Lv_2018} which overlaps ours in a substantial way. They considered two-dimensional droplets lying on a flat substrate, horizontal or tilted. They solved Young-Laplace's equation with a method strongly inspired by the work of Landau and Lifshitz \cite{Landau_1987}. This means that droplets are put in  artificial and unnecessarily complicated settings composed of meniscus walls and holes in solid-liquid interfaces, etc. Since they focus mostly on getting exact solutions and its technicalities, the physics behind hydrostatic droplet configurations was largely neglected, or only briefly mentioned as concluding remarks. And the method they proposed to solve the problem is too complicated to be useful, it has very limited applications. For instance, it could not be applied in  interesting situations like droplets hanging on horizontal or tilted ceilings. In our work we discuss more physics and use more generally applicable method in a minimal setting to tackle the problem. We have compared our result for droplets on a horizontal or inclined plane with Lv and Shi's. We find they agree.

We will start with discussions of friction at a solid-liquid interface. For that we will have to discuss stresses and strains arising at the interface. Obviously the Young's equation is one result we make sure to reinstate in this discussion, though in a modified form. Then we investigate the possible role of static friction in contact angle hysteresis. Formerly, friction has been pointed out as the cause of hysteresis \cite{Joanny_1984, deGennes_2003, Drelich_2020, Adam_Jessop_1925, Butt_2022, Daniel_2023, Good_1952, Tenan_1982, Santos_2011, Nosonovsky_2015, Makkonen_2017, Heijden_2018, Law_2022, YijieXiang_2022, McHale_2022, Zhang_2023, Lindeman_2023, Gao_2018}, but the understanding was not quite complete in our opinion. Here we will provide a very detailed analysis on it.

Consider a lump of solid, crystalline or polycrystalline. It could be placed in vacuum, or in air, with or without contact with another solid substrate. The atoms in the solid lump are to be located at their respective stable equilibrium positions of the total potential energy for a system of a large number of interacting atoms. Atoms deep in the bulk are expected to display a very regular crystal structure, but this is not true for atoms near the surface since they do not see the same distribution of atoms. There is asymmetry inherent from the existence of the surface itself. This asymmetry disappears very rapidly as we move away from the surface into the bulk. Depending on the precise shape of the surface, the structure near the surface would be out of sync from that of the bulk. This is why we have quite different energy levels and hence different electronic properties for nanoparticles for which most atoms cannot be said to be in the bulk. 

The elasticity of a solid is defined by the degree of deviation ('strain') from this equilibrium configuration of atoms when acted by an external force ('stress') on a surface area. It means that the elastic moduli in principle depend on the local geometry and the nature of the interface. In the case of a droplet lying on a substrate we have three different interfaces meeting at a contact line where we expect the force balance. The stress tensor on an isotropic solid interface can be expressed as
\begin{equation*}
    \sigma_{ij} = (\gamma + \lambda w_{\ell\ell}) \delta_{ij} + 2 \mu w_{ij}
\end{equation*}
where $w_{ij}=(1/2)(\partial w_i/\partial x_j + \partial w_j/\partial x_i)$ is the strain tensor ($i,j=1,2$), $\gamma$ is the surface tension, and $\lambda$ and $\mu$ are the 2-dimensional Lam\'e coefficients. ($\gamma$, $\lambda$, and $\mu$ all have the dimension of {\it force/length}\/.) Here we put the surface tension term separately since there is stress left on the surface even if all strains due to the external stresses go to zero. It was inherited in the process of forming the surface, and thus it should be distinguished from the additional stress coming from the external actions. (One can infer the existence of the surface tension term from considering the quasi-static liquidation process of a solid lump for which strain disappears but obviously the surface tension should remain.) The coefficients $\gamma$, $\lambda$, and $\mu$ differ for solid-gas and solid-liquid interfaces. When we demand the lateral force balance at a contact point, the participating forces are these stresses from the surface tensions plus those from any existent external forces like friction which develop strain. Thus the conventional Young's equation is modified to be (here the index 1 representing the lateral direction)
\begin{equation*}
    \sigma^{sg}_{11} - \sigma^{sl}_{11} = \gamma^{lg} \cos \theta_{M}
\end{equation*}
where $\sigma^{sg}_{11}- \sigma^{sl}_{11}= \gamma^{sg} - \gamma^{sl} \pm \kappa$, and $\kappa$ denotes the static friction resisting the added external force. Obviously the static friction enters in the form of the strain ($w_{11}$) multiplied by the elastic moduli $\lambda$ and $\mu$ which generate the opposing stress. Since the strain $w_{11}$ is the rate of lateral change of the lateral deviation, there should be a maximum value beyond which the strain cannot be increased, implying that the contact line would move when there is too strong external force. This is the well-known phenomenon of contact angle hysteresis. At this critical value of strain or the maximum static friction we propose to define the coefficient of static friction. 

For a solid-solid interface the coefficient of static friction is defined as the ratio between the maximum static friction and the normal force between the two solids. In the case of solid-liquid interfacial friction the pinning force at the contact line would be the right physical quantity to play the role of normal force. For a liquid drop lying on a substrate the weight per unit area would be uniformly distributed over the entire contact area, which is irrelevant in pinning a liquid drop. The downward pinning force should be the direct cause of static friction at contact points \cite{Berry_1971, Berry_1975, Tadmor_2021, Marchand_2011a, Marchand_2011b, Fan_2020, Durand_2021}. Thus the maximum static friction is given as the product of the coefficient of static friction and the pinning force at the advancing or the receding angle, whichever breaks the static balance first:
\begin{equation*}
    \kappa_{\mathrm{max}}=|\gamma(\cos\theta_{Y} -\cos\theta_{c})|=\mu_s \gamma\sin\theta_{c}
\end{equation*}
where we put $\Delta\gamma \equiv \gamma^{sg} - \gamma^{sl}=\gamma^{lg}\cos\theta_{Y}$ and $\gamma \equiv \gamma^{lg}$ for convenience, and $\theta_{c}$ is either the advancing contact angle $\theta_{a}$ or the receding contact angle $\theta_{r}$. Here we are making the simplest possible choice of linearity between friction and pinning force, but we might have to confront other possibilities, which will not be pursued here. If the coefficient of static friction $\mu_s$ is known, by solving the above equation we can determine the critical contact angle $\theta_{c}$ (which is either $\theta_{a}$ or $\theta_{r}$) as in the following
\begin{equation*}
    \theta_{c} = \cos^{-1} \left(\frac{\cos \theta_{Y}}{\sqrt{1 + \mu_s^2}}\right) \pm \tan^{-1} \mu_s \ .
\end{equation*}

Contact angle hysteresis can be observed everywhere. For instance, look at a water droplet put on a horizontal glass plate, and see what happens when we tilt the plate very slowly. The contact angle at the bottom of the droplet would increase, but at the top it would decrease, hence the overall shape of the droplet would change. This will go on until one of the contact points moves. Certainly the precise values of advancing and receding contact angles would depend on the Bond number and the properties and conditions of the water-glass interface. For relatively small droplets one can even complete a full circle of the glass plate during which an initial sessile droplet turns into a pendent one and then back to the original sessile one. In our work we can understand this problem in a very thorough and systematic way at least for two-dimensional droplets. 

The structure of this work is as follows. In Section II we set up the problem and derive the governing equations and the constraints. We solve the equations numerically for various values of inclination angle and Bond number. We show the changing profiles of droplets on an inclined plane when we make a half circle of rotation of the plane: from the sessiles to the pendents. In Section III the coefficient of static friction is defined and is expressed in terms of the critical angles. For comparison, in Section IV, we consider the case of droplets when we ignore the gravity. And in the ensuing sections we study two exactly solvable cases of sessile droplets (Section V) and pendent droplets (Section VI) in the presence of gravity. In Section VII we will find there could be two stable profiles for a single Bond number, which we dubbed the bifurcation. In the penultimate section (VIII) we discuss the possible pinch-offs for pendent droplets and find a simple relation between the fallen-out volume and the Bond number at the pinch-off. In the last section (IX) we will summarize our results and conclude.

\section{A droplet on an inclined plane}

\begin{figure}[!t]
    \includegraphics[width=0.45\textwidth]{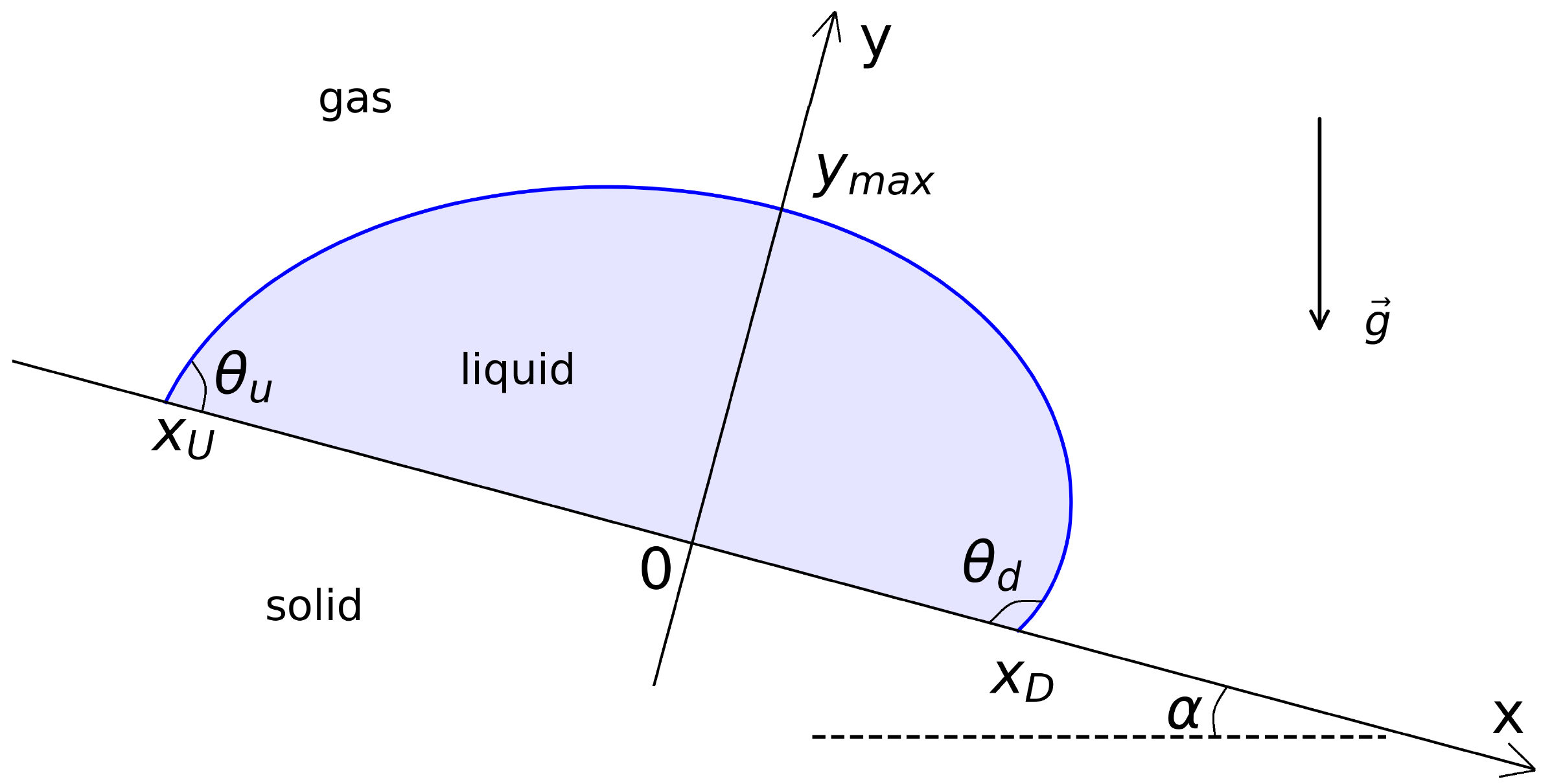}
    \caption{A generic shape of a 2-dimensional droplet on an inclined plane.}
    \label{Figure1}
\end{figure}

Consider a 2-dimensional droplet stuck on an inclined planar substrate. As we see in Fig.\ref{Figure1}, the shapes of droplets differ by several factors: total cross-sectional area, total contact length, contact angles, inclination angle, and so on. The conditions and the equations that govern the profile of a static droplet of cross-sectional area $A$ are given as follows: First, the profile of a droplet is related to its area $A$ by
\begin{equation*}
    \int_{0}^{y_{m}} \left[x_{d} \left(y\right) - x_{u} \left(y\right)\right] \,dy = A
\end{equation*}
where $x_d(y)$ and $x_u(y)$ are the profile functions for $x \geq 0$ and $x \leq 0$, respectively. Second, since we will assume the contact points are fixed due to static frictions ({\it hysteresis}), we have
\begin{equation*}
    x_{u} \left(0\right) = x_{U}, \qquad x_{d} \left(0\right) = x_{D}
\end{equation*}
where $x_D - x_U \equiv 2 x_0$ and the total contact length $2x_0$ is a fixed parameter. And last, the profile of the droplet will be determined by minimizing its energy ({\it energy/length} more precisely) while satisfying the constraints in the above.
\begin{widetext}
\begin{align*}
    \begin{split}
        \mathcal{E} = &\gamma \int_{0}^{y_{m}} \sqrt{1 + \left\{x_{d}' \left(y\right)\right\}^{2}} \,dy + \rho g \int_{0}^{y_{m}} \left[x_{d} \left(y\right) y \cos \alpha - \frac{1}{2} x_{d}^{2} \left(y\right) \sin \alpha\right] \,dy - \Delta \gamma x_{d} \left(0\right)\\
                    + &\gamma \int_{0}^{y_{m}} \sqrt{1 + \left\{x_{u}' \left(y\right)\right\}^{2}} \,dy - \rho g \int_{0}^{y_{m}} \left[x_{u} \left(y\right) y \cos \alpha - \frac{1}{2} x_{u}^{2} \left(y\right) \sin \alpha\right] \,dy + \Delta \gamma x_{u} \left(0\right)\\
                    - &\beta \left[\int_{0}^{y_{m}} \left[x_{d} \left(y\right) - x_{u} \left(y\right)\right] \,dy - A\right] + \kappa_{d} \left[x_{d} \left(0\right) - x_{D}\right] + \kappa_{u} \left[x_{u} \left(0\right) - x_{U}\right]
    \end{split}
\end{align*}
\end{widetext}
where we introduced $\beta$, $\kappa_{d}$, $\kappa_{u}$ as Lagrange multipliers for three constraints, and $y_{m} \equiv y_{\mathrm{max}}$.

If we minimize the above energy for the variations of droplet profile, we can obtain the profile equations for $x_u(y)$ and $x_d(y)$ as well as the constraint equations \cite{Shanahan_1982, Shanahan_1984, Safran_1994, Carre_1995, Bullard_2005, Heijden_2018, Shkolnikov_2019}:

At contact points we have the modified Young's equations
\begin{equation*}
    \gamma \cos \theta_{d} - \Delta \gamma + \kappa_{d} = 0, \qquad \gamma \cos \theta_{u} - \Delta \gamma - \kappa_{u} = 0 \ .
\end{equation*}
We see that the Lagrange multipliers $\kappa_{u}$, $\kappa_{d}$ which are the constraint forces can be interpreted as static frictions originating from the pinning forces at contact points \cite{Adam_Jessop_1925, Good_1952, Extrand_Gent_1990, Carre_1995, Santos_2011, Makkonen_2017, Heijden_2018, Fan_2020, Odunsi_2023}. Since $\Delta \gamma = \gamma \cos \theta_{Y}$, the static friction at each contact point can be re-expressed as 
\begin{equation*}
    \kappa_{d} = \gamma \left(\cos \theta_{Y} - \cos \theta_{d}\right), \qquad \kappa_{u} = \gamma \left(\cos \theta_{u} - \cos \theta_{Y}\right) \ .
\end{equation*}

The profile equations are the well-known Young-Laplace's equations
\begin{align} \label{eqn: Young-Laplace_eq_ud}
    \begin{split}
        - \gamma \frac{d}{dy} \left(\frac{x_{d}' \left(y\right)}{\sqrt{1 + \left(x_{d}'\right)^{2}}}\right) &+ y \rho g \cos \alpha \\
                                                                                                            &- x_{d} \left(y\right) \rho g \sin \alpha - \beta = 0 \\
        \\
        - \gamma \frac{d}{dy} \left(\frac{x_{u}' \left(y\right)}{\sqrt{1 + \left(x_{u}'\right)^{2}}}\right) &- y \rho g \cos \alpha \\
                                                                                                            &+ x_{u} \left(y\right) \rho g \sin \alpha + \beta = 0 \ .
    \end{split}
\end{align}

\begin{figure*}[!t]
  \includegraphics[width=1.0\textwidth]{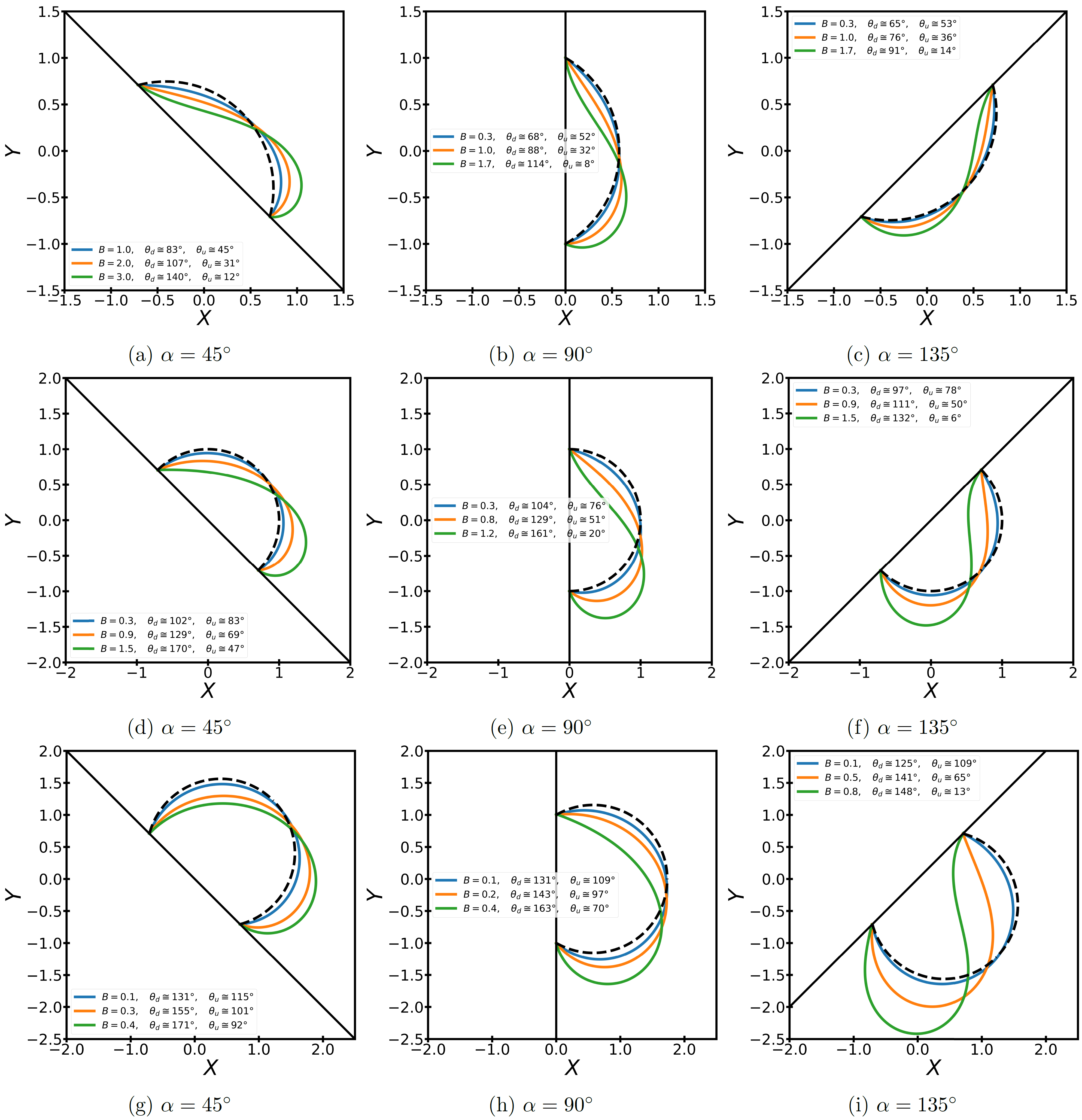}
  \caption{Profiles of 2-dimensional droplets on an inclined plane for various Bond numbers and inclination angles.
           (a)-(c): $\theta_{C} = 60^{\circ}$, (d)-(f): $\theta_{C} = 90^{\circ}$, and (g)-(j): $\theta_{C} = 120^{\circ}$.
           The black dashed lines are for $B = 0$ ($\theta_{d} = \theta_{u} = \theta_{C}$).}
  \label{Figure2}
\end{figure*}

By integrating the Young-Laplace's equations Eq.(\ref{eqn: Young-Laplace_eq_ud}) in the interval $\left[0, y_{m}\right]$ we can obtain the relations for the remaining Lagrange multiplier $\beta$:
\begin{align} \label{eqn: beta_i_down}
  \begin{split}
    \beta = \gamma \frac{1 - \cos \theta_{d}}{y_{m}} &+ \frac{1}{2} y_{m} \rho g \cos \alpha \\
                                                     &- \frac{1}{y_{m}} \rho g \sin \alpha \int_{0}^{y_{m}} x_{d} \left(y\right) \,dy
  \end{split}
\end{align}
and
\begin{align} \label{eqn: beta_i_up}
  \begin{split}
    \beta = \gamma \frac{1 - \cos \theta_{u}}{y_{m}} &+ \frac{1}{2} y_{m} \rho g \cos \alpha \\
                                                     &- \frac{1}{y_{m}} \rho g \sin \alpha \int_{0}^{y_{m}} x_{u} \left(y\right) \,dy \ .
  \end{split}
\end{align}

\begin{figure*}[!t]
  \includegraphics[width=0.9\textwidth]{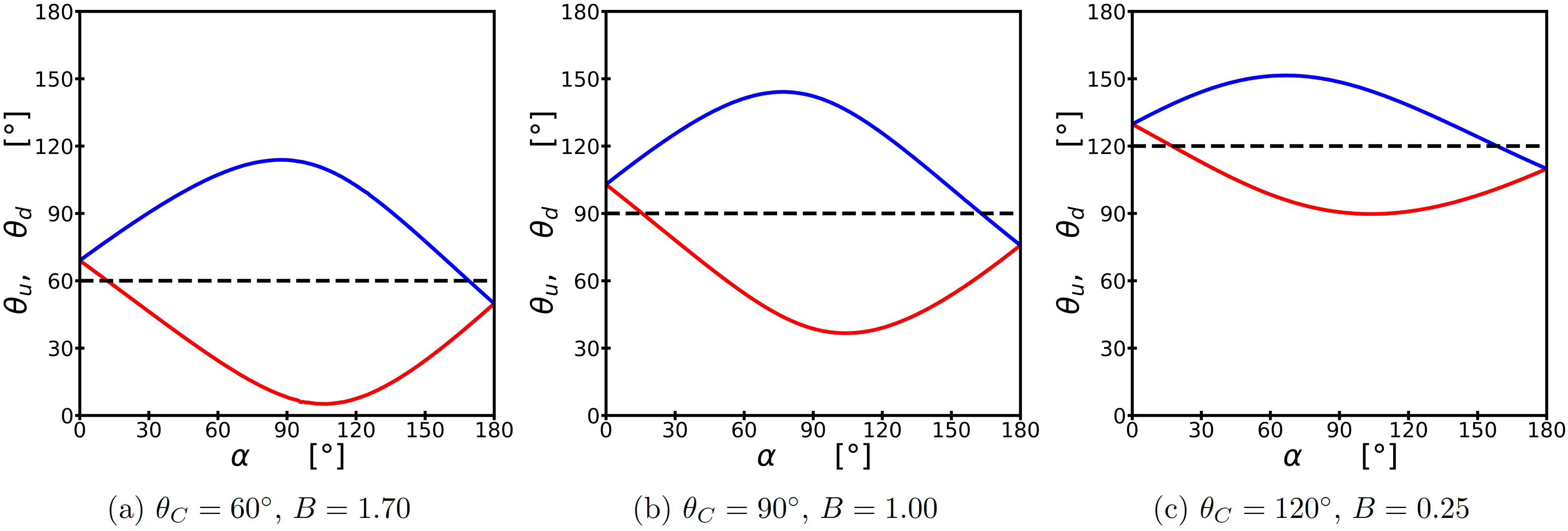}
  \caption{Changes in the contact angles relative to the inclination angle. We plotted $\theta_{d}$ (blue), $\theta_{u}$ (red), and $\theta_{C}$ (black dashed).}
  \label{Figure3}
\end{figure*}

\noindent Adding these equations, $\beta$ is determined as
\begin{equation*}
    \beta = \gamma \frac{2 - \cos \theta_{d} - \cos \theta_{u}}{2 y_{m}} + \frac{1}{2} y_{m} \rho g \cos \alpha - \frac{1}{2} \frac{\Delta A}{y_{m}} \rho g \sin \alpha
\end{equation*}
where 
$\Delta A$ is the difference in areas
\begin{equation*}
    \Delta A \equiv \int_{0}^{y_{m}} \left[x_{d} \left(y\right) + x_{u} \left(y\right)\right] \,dy = A_{d} - A_{u}
\end{equation*}
and
\begin{equation*}
    A_{d} \equiv \int_{0}^{y_{m}} x_{d} \left(y\right) \,dy, \qquad A_{u} \equiv - \int_{0}^{y_{m}} x_{u} \left(y\right) \,dy \ .
\end{equation*}
We can interpret $\beta$ as the total normal force per unit length acting on the droplet by the substrate. 

Subtracting Eq.(\ref{eqn: beta_i_down}) and Eq.(\ref{eqn: beta_i_up}), we get another useful relation
\begin{equation*}
    A \rho g \sin \alpha = \gamma \left(\cos \theta_{u} - \cos \theta_{d}\right)=\kappa_{d} + \kappa_{u}
\end{equation*}
which is none other than the force balance equation applied to the whole droplet, not just the balance equation at each contact point. Observe that $\theta_{u} \leq \theta_{d}$. The so-called Furmidge relation \cite{Macdougall_1942, Furmidge_1962, Dussan_1983, Carre_1995, Extrand_1995, Krasovitski_2005, Bouteau_2008, Tadmor_2008, Gao_2018, Shkolnikov_2019, Butt_2022} can be obtained by substituting the advancing angle $\theta_{a}$ for $\theta_{d}$ and the receding angle $\theta_{r}$ for $\theta_{u}$. However, we think that these substitutions are unjustified since the receding angle and the advancing angle need not always be simultaneously attained, which was already pointed out in previous works \cite{White_2015, Butt_2022, Krasovitski_2005, Santos_2011, Law_2022, Oner_2000}.

To solve the Young-Laplace's equations we first introduce dimensionless variables and quantities such as $X \equiv x / x_{0}$, $Y \equiv y / x_{0}$, $A^{*} \equiv A / x_{0}^{2}$, $\beta^{*} \equiv \beta x_{0} / \gamma$, and $B \equiv \rho g x_{0}^{2} / \gamma$.

Now the dimensionless Young-Laplace's equations are
\begin{align} \label{eqn: Young-Laplace_eq_ud_1}
    \begin{split}
        - \frac{d}{dY} \left(\frac{X_{d}' \left(Y\right)}{\sqrt{1 + \left(X_{d}'\right)^{2}}}\right) &+ Y B \cos \alpha \\
                                                                                                     &- X_{d} \left(Y\right) B \sin \alpha - \beta^{*} = 0 \\
        - \frac{d}{dY} \left(\frac{X_{u}' \left(Y\right)}{\sqrt{1 + \left(X_{u}'\right)^{2}}}\right) &- Y B \cos \alpha \\
                                                                                                     &+ X_{u} \left(Y\right) B \sin \alpha + \beta^{*} = 0
    \end{split}
\end{align}
where
\begin{equation*}
    \beta^{*} = \frac{2 - \cos \theta_{d} - \cos \theta_{u}}{2 Y_{m}} + \frac{1}{2} Y_{m} B \cos \alpha - \frac{1}{2} \frac{\Delta A^{*}}{Y_{m}} B \sin \alpha \ .
\end{equation*}
Analytical solutions to Eq.(\ref{eqn: Young-Laplace_eq_ud_1}) do not exist except for the cases of special inclination angles $\alpha=0^{\circ}$, $90^{\circ}$, and $180^{\circ}$. We use numerical methods to obtain the solutions for a general inclination angle. In Fig.\ref{Figure2} we have displayed the profiles of droplets for a variety of inclination angles ($\alpha$), contact angles without gravity ($\theta_{C}$, see Section IV), and Bond numbers ($B$). We also show the up- and down-contact angles for some appropriately chosen values of $\theta_{C}$ and $B$ as functions of $\alpha$ in Fig.\ref{Figure3}. 

Some experimental and numerical works on contact angle hysteresis have been done for 3-dimensional droplets \cite{Brown_1980, Shanahan_1982, Shanahan_1984, Milinazzo_1988, Bouteau_2008, Santos_2011, Santos_2012, Berthier_2012, White_2015, Pozrikidis_2016, Yonemoto_2017, JinYoungKim_2017, Timm_2019, Hattori_2019, Coninck_2021, Yonemoto_2022}. Their conclusions are qualitatively in good agreement with ours even though our works are concerned with 2-dimensional droplets only: Contact lines do not budge but contact angles change as Bond number or inclination angle change. Either advancing contact line or receding contact line first starts to move, but not simultaneously. Even the cross-sectional views of 3-dimensional droplets are strikingly similar to the shapes of our 2-dimensional droplets \cite{Brown_1980, Shanahan_1982, Shanahan_1984, Milinazzo_1988, Bouteau_2008, Berthier_2012, Pozrikidis_2016, JinYoungKim_2017, Yonemoto_2017, Timm_2019, Coninck_2021}.

\section{Coefficient of static friction and critical contact angle}

\begin{figure*}[!t]
  \includegraphics[width=0.9\textwidth]{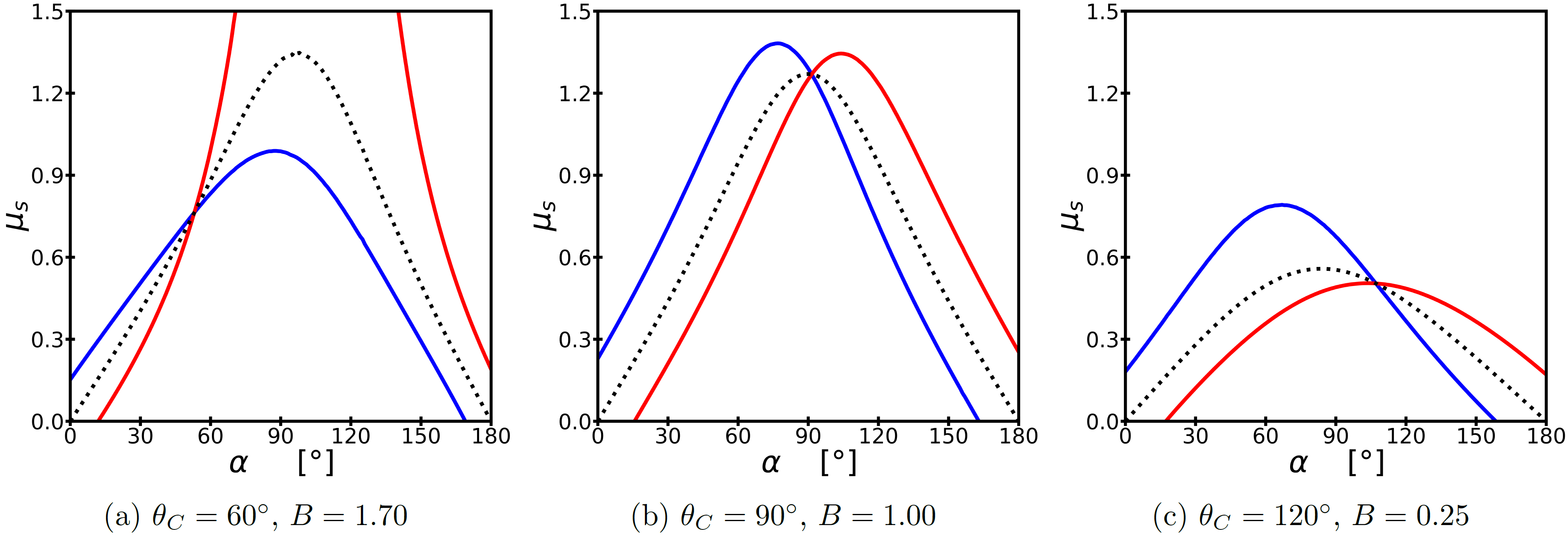}
  \caption{The blue solid lines are for $\theta_{c} = \theta_{a} = \theta_{d}$, the red lines for $\theta_{c} = \theta_{r} = \theta_{u}$, and the black dotted lines for $\theta_{d} = \theta_{a}$, $\theta_{u} = \theta_{r}$.
           For given $\theta_{C}$ and $B$ a droplet becomes stable for any inclination angle once $\mu_{s}$ assumes a value above all three lines.}
  \label{Figure4}
\end{figure*}

Taking analogy with the solid-solid friction we define the coefficient of static friction as the ratio between the static friction (constraint force) at the critical angle, advancing or receding, and the pinning force \cite{McHale_2022, Zhang_2023, Gao_2018}.
\begin{equation*}
\mu_s\equiv \frac{\left|\cos \theta_{Y} - \cos \theta_{c}\right|}{\sin \theta_{c}} 
\end{equation*}
where $\theta_{c} = \theta_{a} > \theta_{Y}$ or $\theta_{r} < \theta_{Y}$ whichever comes first to move the contact point \cite{White_2015, Butt_2022, Krasovitski_2005, Santos_2011, Law_2022, Oner_2000}. Obviously the friction coefficient, thus defined, depends on the inclination angle of the substrate plane, which sounds odd since the coefficient of friction should reflect the inherent nature of interfaces, not the geometric setting like inclination angle. However, this is inevitable since the critical angles are dependent on various geometric factors like inclination angle. Thus, the usefulness of this definition would be debatable in general situations. Controlling one or two parameters in a restricted setting would render our definition and interpretation the power to analyze the problem thoroughly.

There are three distinct cases of critical contact angle attained either (1) at the bottom side first ($\theta_{c} = \theta_{a} = \theta_{d}$, $\theta_{u} > \theta_{r}$), (2) at the top side first ($\theta_{c} = \theta_{r} = \theta_{u}$, $\theta_{d} < \theta_{a}$), or (3) simultaneously ($\theta_{d} = \theta_{a}$, $\theta_{u} = \theta_{r}$). The coefficient $\mu_s$ for three different cases assumes the following expressions:
\begin{equation*} \label{eqn: mu}
    \mu_{s} =
    \begin{cases} 
        \dfrac{\cos \theta_{Y} - \cos \theta_{d}}{\sin \theta_{d}} & \qquad \left(\theta_{c} = \theta_{a} = \theta_{d}, \quad \theta_{u} > \theta_{r}\right) \\
        \\
        \dfrac{\cos \theta_{u} - \cos \theta_{Y}}{\sin \theta_{u}} & \qquad \left(\theta_{c} = \theta_{r} = \theta_{u}, \quad \theta_{d} < \theta_{a}\right) \\
        \\
        \tan \left(\dfrac{\theta_{d} - \theta_{u}}{2}\right)                    & \qquad \left(\theta_{d} = \theta_{a}, \quad \theta_{u} = \theta_{r}\right) \ .
    \end{cases}
\end{equation*}
The last case of the above equation has been discussed in McHale et al. \cite{McHale_2022}. From our perspective, however, their work applies only for a very narrow class of liquids and solid substrates. For illustration, in Fig.\ref{Figure4} the coefficients of static friction are plotted against inclination angle for three different initial Young's angles (choosing $\theta_{C} = \theta_{Y}$ for convenience) and Bond numbers.

It is possible to determine the coefficient of static friction as well as the advancing or the receding contact angles by varying the inclination angle adiabatically and measuring the contact angle just before the liquid droplet starts to move.

\section{A droplet without gravity}
When we ignore the presence of gravity, which is approximately true for very small droplets, the profile equations can be solved exactly. Obviously the profiles are circular, but they should be compatible with the constraints, i.e. fixed area and fixed contact points. Substituting $B = 0$ into the equations derived in Section II, we can get the profile equations for a 2-dimensional droplet without gravity.

We have the following force balance equation:
\begin{equation}
    \kappa_{C} = \gamma \left(\cos \theta_{Y} - \cos \theta_{C}\right) \neq 0 \label{eqn: Young_Eq_c}
\end{equation}
where $\theta_{C}$ is the contact angle when the profile is a part of a circle. Though we are neglecting the gravity, there still exists static friction, hence contact angle hysteresis. The contact angle must be modified and is different from the Young's angle. In other words, the contact angle $\theta_{C}$ need not be fixed as $\theta_{Y}$ and it is only required to be in the range between the receding contact angle and the advancing contact angle. Without taking into account the contact angle hysteresis due to static friction, the contact angle should be equated to the Young's angle and hence the contact points should be adjusted accordingly to achieve a static configuration. 

For an actual droplet on a substrate of complicated topography finding and settling at a stable equilibrium looks like a very difficult process. In contrast to this the contact angle hysteresis due to static friction naturally explains why a liquid drop easily settles into an equilibrium configuration. We should note that it is very difficult to separate out and measure the Young's angle unambiguously in an experimental setting due to the existence of contact angle hysteresis.

\begin{figure}[!b]
    \includegraphics[width=0.45\textwidth]{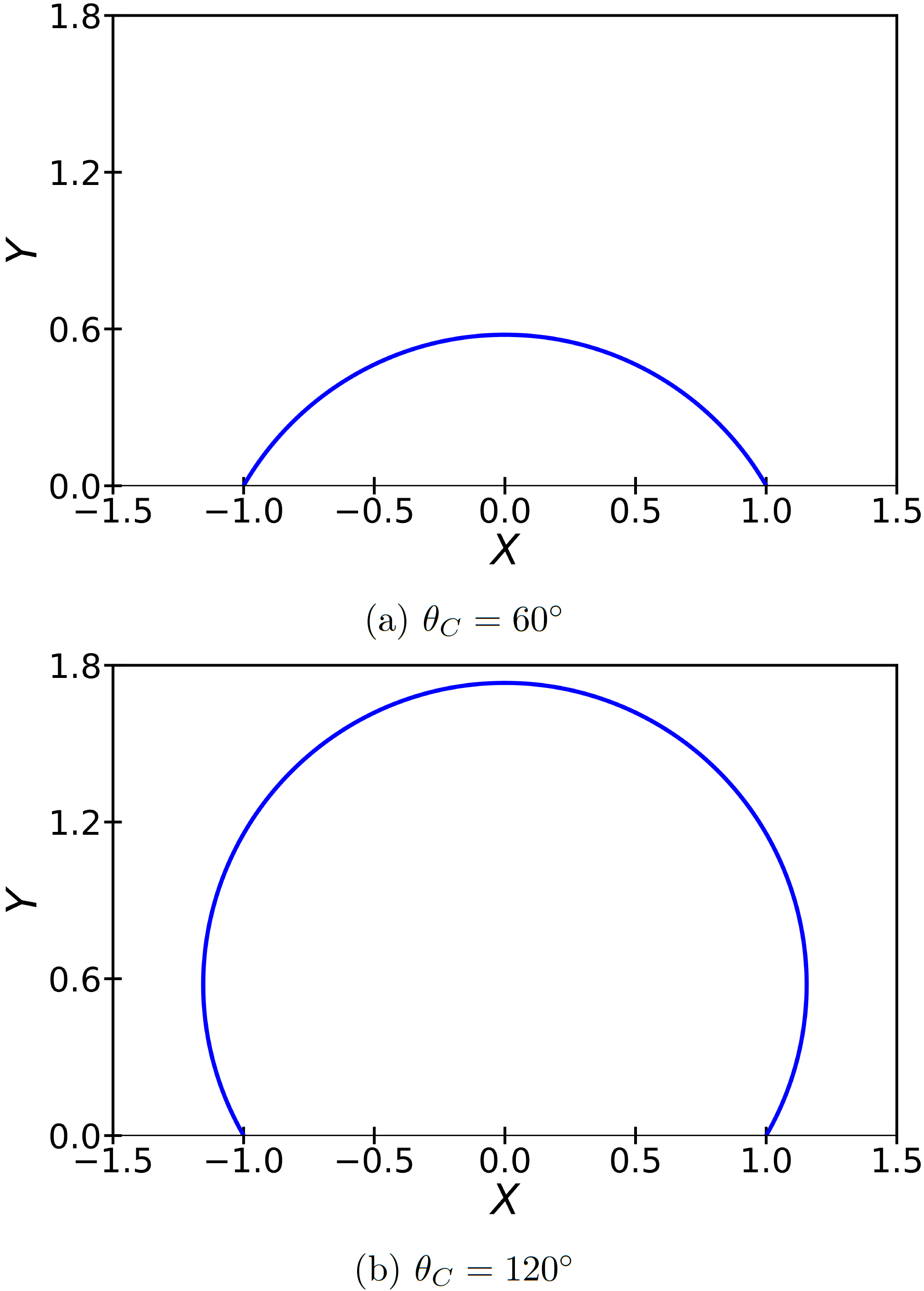}
    \caption{Profiles of 2-dimensional droplets when we ignore gravity.}
    \label{Figure5}
\end{figure}

The circular profile, $X_{C} \left(Y\right)$, can be obtained by integrating the Young-Laplace's equation directly. Fig.\ref{Figure5} shows the shape of 2-dimensional droplets without gravity when the contact angle is $\theta_{C} = 60^{\circ}$ and $120^{\circ}$.
\begin{equation*}
    X_{C} \left(Y\right) = \sqrt{\left(R^{*}\right)^{2} - \left(Y+ R^{*} \cos \theta_{C}\right)^{2}}
\end{equation*}
Here $R^{*}$ is the dimensionless radius which is related to the contact angle $\theta_{C}$ as $R^{*} = 1 / \sin \theta_{C}$.
$R^{*}$ is also related to the Lagrange multiplier $\beta_{C}^{*}$ as $\beta_{C}^{*} = 1 / R^{*}$.
And the dimensionless area $A^{*}$ is determined from the contact angle as
\begin{equation*}
    A^{*} = \frac{\theta_{C} - \sin \theta_{C} \cos \theta_{C}}{\sin^{2} \theta_{C}} \ .
\end{equation*}

\section{Sessile droplets with gravity}
Let us consider a sessile droplet in the presence of gravity: $B > 0$ and $\alpha = 0^{\circ}$. Due to the gravity, the droplet is not of a circular profile any more. Instead of using the generic suffixes of Section II, we will use '$s$' to emphasize we are concerned with 'sessile droplets'. The gravity makes the maximum height ($Y_{s}$) decrease and does the contact angle ($\theta_{s}$) increase. This should be the case in order to have the constraints of fixed contact points and fixed area to be satisfied. Note that $Y_{s} \leq Y_{C}$ and $\theta_{s} \geq \theta_{C}$. $\theta_{s}$ will be dubbed the 'sessile angle'.

\begin{figure}[!b]
  \includegraphics[width=0.45\textwidth]{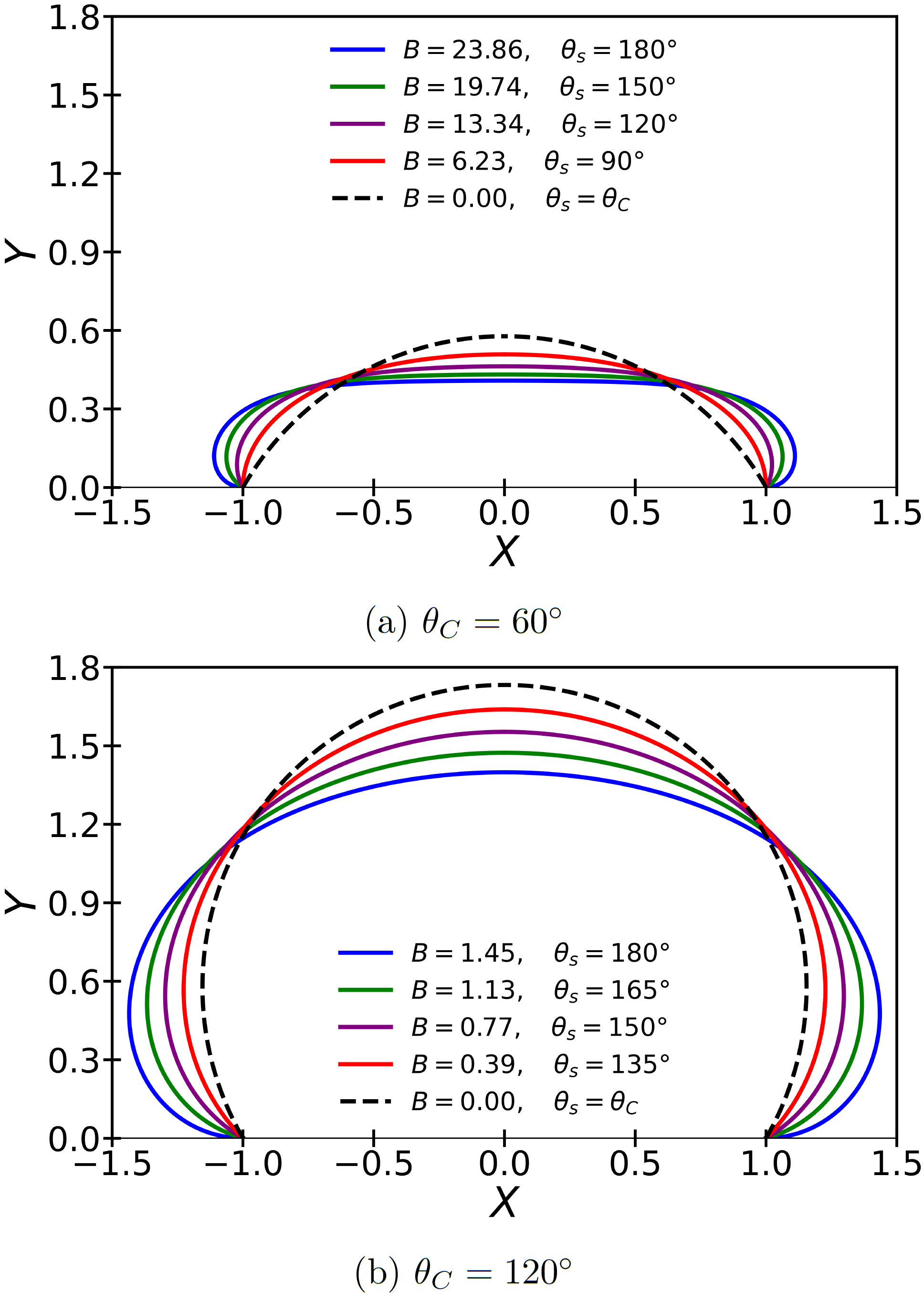}
  \caption{Profiles of 2-dimensional sessile droplets under gravity.}
  \label{Figure6}
\end{figure}

Substituting $\alpha = 0^{\circ}$ into the equations of Section II, the static friction in the modified Young's equation is given as
\begin{equation*}
    \kappa_{s} = \gamma \left(\cos \theta_{Y} - \cos \theta_{s}\right) \ .
\end{equation*}
Combining with Eq.(\ref{eqn: Young_Eq_c}),
\begin{equation*}
    \kappa_{s} - \kappa_{C} = \gamma \left(\cos \theta_{C} - \cos \theta_{s}\right) \ .
\end{equation*}
The sessile angle is always greater than or equal to $\theta_{C}$, so $\kappa_{s} \geq \kappa_{C}$. This means static friction should act on the contact points to prevent the total contact length from increasing. The sessile angle cannot increase indefinitely, hence there is a critical value called the advancing contact angle $\theta_{a}$. In terms of the coefficient of static friction $\mu_{s}$, it is
\begin{equation*}
    \theta_{a} = \cos ^{-1} \left(\frac{\cos \theta_{Y}}{\sqrt{1 + \mu_{s}^{2}}}\right) + \tan^{-1} \mu_{s} \ .
\end{equation*}
Note that $\theta_{a} \rightarrow 180^{\circ}$ as $\mu_{s} \rightarrow \infty$. 

To find the profile of a sessile droplet, we should solve the equation
\begin{equation}
     X_{s}'\left(Y\right) = \frac{f_{s}\left(Y\right)}{\sqrt{1 - \left\{f_{s}\left(Y\right)\right\}^{2}}} \label{eqn: sessile_ODE}
\end{equation}
where
\begin{equation*}
    f_{s}\left(Y\right) = \frac{1}{2} B \left(\bar{Y}_{s} - Y\right)^{2} + \bar{f}_{s},
\end{equation*}
\begin{equation*}
    \bar{Y}_{s} = \frac{1}{2} A^{*} + \frac{\sin \theta_{s}}{B}, \qquad \bar{f}_{s} = - \frac{1}{2} B \bar{Y}_{s}^{2} - \cos \theta_{s} \ .
\end{equation*}
We also have a relation
\begin{equation}
    B Y_{s}^{2} - \left(A^{*} B + 2 \sin \theta_{s}\right) Y_{s} + \left(2 - 2 \cos \theta_{s}\right) = 0 \label{eqn: sessile_condition_2}
\end{equation}

whose solution is 
\begin{equation}
    Y_{s} = \bar{Y}_{s} - \sqrt{\bar{Y}_{s}^{2} - \frac{2}{B} \left(1 - \cos \theta_{s}\right)} \ . \label{eqn: Y_s}
\end{equation}
Since $X_{s}''\left(Y\right) = f_{s}' \left(Y\right) \left[1 - \left\{f_{s} \left(Y\right)\right\}^{2}\right]^{-3/2}$, $Y = \bar{Y}_{s}$ is the inflection point of the sessile droplet. However, since $Y_{s} <\bar{Y}_{s}$, there is no inflection point in the profile of a sessile droplet. This is in contrast to the case of pendent droplet which will be the subject of next section.

We can get the exact profile of a sessile droplet, $X_{s} \left(Y\right)$, by integrating Eq.(\ref{eqn: sessile_ODE}) in terms of elliptic functions \cite{Dwight_1961}:
\begin{align} \label{eqn: sessile_sol}
  \begin{split}
    X_{s} \left(Y\right) = \frac{1}{\sqrt{B}} &\left[\frac{2}{k} \left\{E\left(\phi, k\right) - E\left(\frac{\pi}{2}, k\right)\right\}\right. \\
                                              &\left.+ \frac{k^{2} - 2}{k} \left\{F\left(\phi, k\right) - F\left(\frac{\pi}{2}, k\right)\right\}\right] \ .
  \end{split}
\end{align}
The boundary condition at $X_{s} \left(0\right) = 1$ implies
\begin{align} \label{eqn: sessile_2}
  \begin{split}
    \sqrt{B} = &\frac{2}{k} \left\{E\left(\phi_{0}, k\right) - E\left(\frac{\pi}{2}, k\right)\right\} \\
               &+ \frac{k^{2} - 2}{k} \left\{F\left(\phi_{0}, k\right) - F\left(\frac{\pi}{2}, k\right)\right\}
  \end{split}
\end{align}
where $k = \sqrt{2 / \left(1 - \bar{f}_{s}\right)}$ and
\begin{equation*}
    \phi = \sin^{-1} \left(\frac{\sqrt{1 - f_{s}\left(Y\right)}}{\sqrt{2}}\right), \qquad \phi_{0} = \frac{\pi}{2} - \frac{\theta_{s}}{2} \ .
\end{equation*}
Here, $\bar{f}_{s}$, $k$, and $\phi_{0}$ are all functions of $\theta_{s}$ only, so we can determine the maximum height $Y_{s}$ and the sessile angle $\theta_{s}$ for given values of $B$ and $\theta_{C}$ from the above relations between $Y_{s}$ and $\theta_{s}$ (Eq.(\ref{eqn: Y_s}), (\ref{eqn: sessile_2})). A few profiles of the 2-dimensional sessile droplet in the presence of gravity for $\theta_{C} = 60^{\circ}$ and $120^{\circ}$ in Eq.(\ref{eqn: sessile_sol}) are shown in Fig.\ref{Figure6}.

\begin{figure}[!t]
    \includegraphics[width=0.45\textwidth]{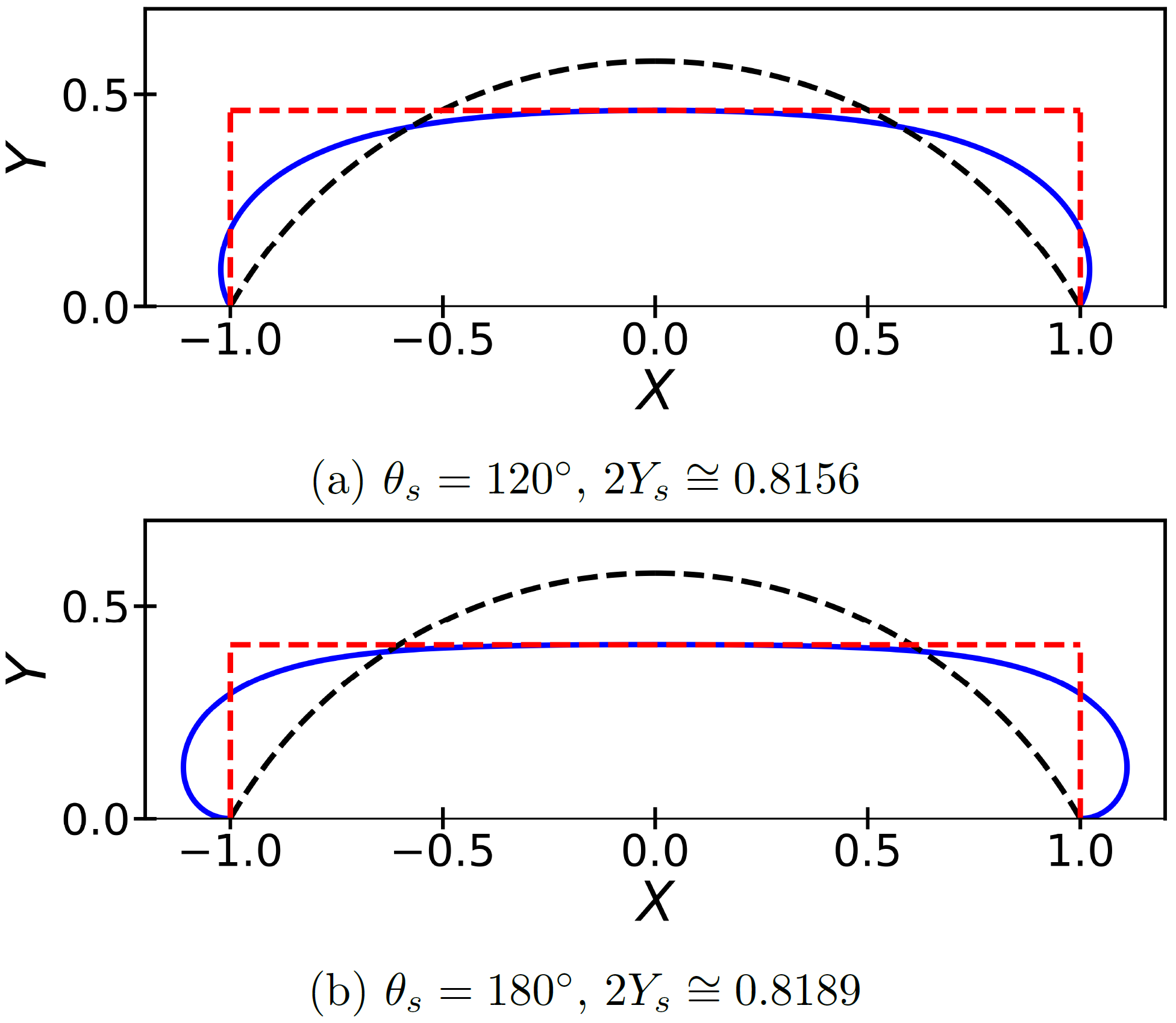}
    \caption{Profiles of sessile droplets in the limit of large $B$ (blue solid lines). In this limit it can be approximated as red dashed rectangles. The black dashed lines are profiles in the absence of gravity.
            Here $\theta_{C} = 60^{\circ}$ and $A^{*} \cong 0.8189$.}
    \label{Figure7}
\end{figure}

\begin{figure*}[!t]
  \includegraphics[width=0.9\textwidth]{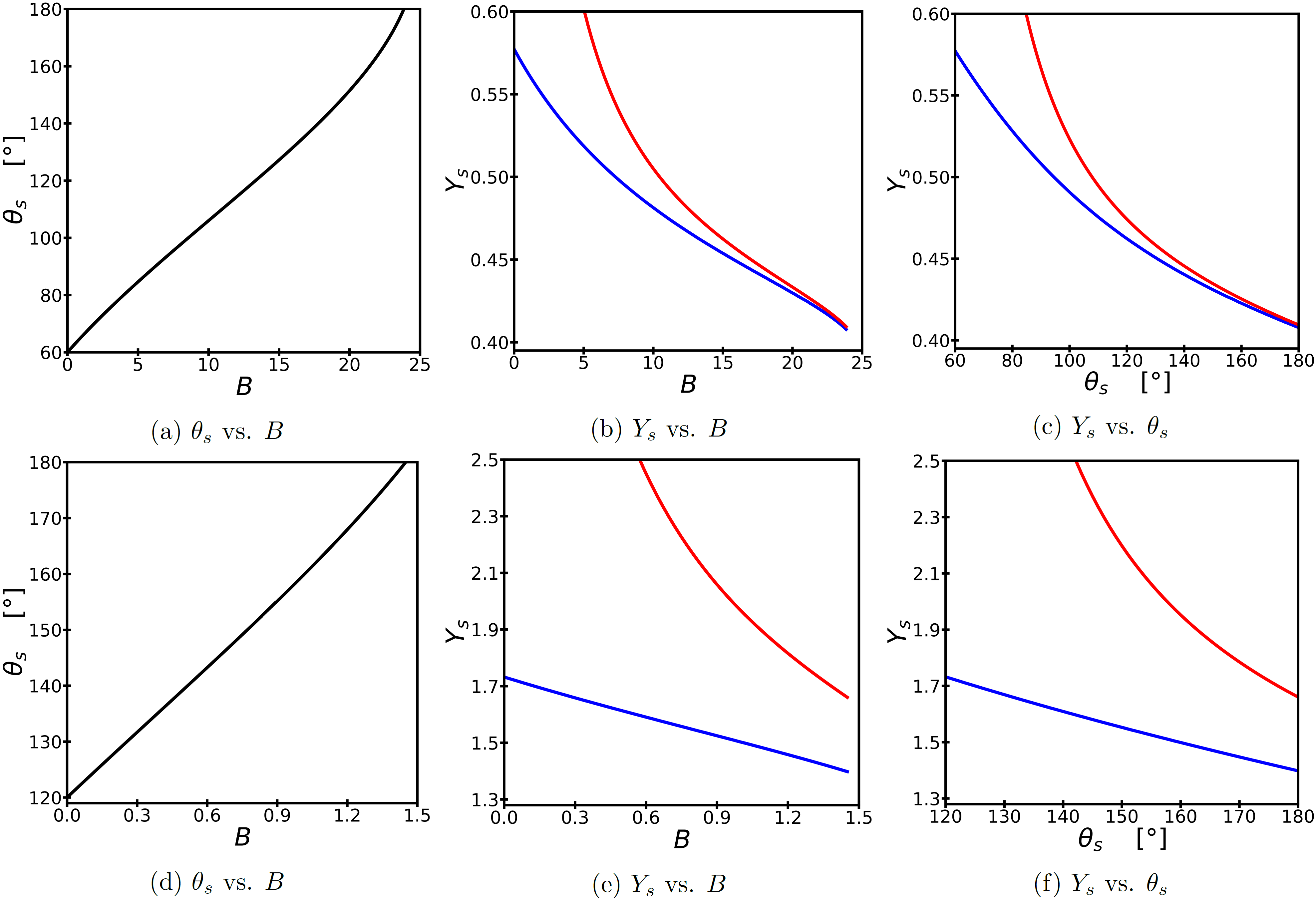}
  \caption{(a)-(c): $\theta_{C} = 60^{\circ}$, (d)-(f): $\theta_{C} = 120^{\circ}$. 
           Blue solid lines are exact solutions, and red solid lines are approximate expression $Y_{s} \cong \left(2 / \sqrt{B}\right) \sin \left(\theta_{s} / 2\right)$.}
  \label{Figure8}
\end{figure*}

We can check the consistency of our solutions by considering a special limit of large Bond number. If gravity is sufficiently large compared to a liquid-gas surface tension, a sessile droplet is expected to become flat. As shown in Fig.\ref{Figure7}, indeed the curve around the maximum height of a droplet becomes flat. In this limit the droplets become 'pancake-like' \cite{Myshkis_1987, Safran_1994, deGennes_2003, Berthier_2012, Shkolnikov_2019, Yariv_2023}. The area of the two-dimensional 'pancake-like' droplet can be approximated as the area of a rectangle with width $2 X_{0} = 2$ and height $Y_{s}$, which is $A^{*} \cong 2 Y_{s}$. In the limit of large $B$ Eq.(\ref{eqn: sessile_condition_2}) is approximated as
\begin{equation*}
    \frac{1}{2} B Y_{s}^{2} + \sin \theta_{s} Y_{s} - \left(1 - \cos \theta_{s}\right) \cong 0
\end{equation*}
whose solution is 
\begin{equation*}
    Y_{s} \cong \frac{2}{\sqrt{B}} \sin \left(\frac{\theta_{s}}{2}\right) \ .
\end{equation*}
It can be seen that the above approximation is correct in the large Bond number limit as compared in Figs.\ref{Figure7} and \ref{Figure8}.

\section{Pendent droplets with gravity}
Now consider a pendent droplet. It is a liquid droplet hung under a planar solid substrate. This corresponds to our parameters having the values, $B \geq 0$ and $\alpha = 180^{\circ}$. (Again, we will use the suffix '$p$' to emphasize all the values are for pendent droplets.) The gravity is downward, but there is the combined effect of pressure difference and surface tension counterbalancing the gravity. In particular, the normal force at contact points ('{\it pinning force}') due to the interfacial tensions always acts in the direction from liquid to solid, i.e. upward for a pendent droplet. This pinning force not only prevents the pendent droplet falling from the ceiling, but also creates the static friction at the contact points, like in the case of sessile droplets. Hence we should expect the contact angle hysteresis.

The static friction on the pendent droplet, $\kappa_{p}$, becomes
\begin{align*}
    &\kappa_{p} = \gamma \left(\cos \theta_{Y} - \cos \theta_{p}\right), \\
    &\kappa_{p} - \kappa_{C} = \gamma \left(\cos \theta_{C} - \cos \theta_{p}\right)
\end{align*}
where the contact angle $\theta_{p}$ will be named as the 'pendent angle'. Note that $\theta_{p} \leq \theta_{C}$. Thus $\kappa_{p} \leq \kappa_{C}$. Hence the direction of the static friction is opposite to that of the sessile droplet.

The pendent angle decreases when the influence of gravity on the pendent droplet increases, so the critical angle is the minimum angle that it can have, in other words, the receding contact angle $\theta_{r}$. In terms of the coefficient of static friction $\mu_{s}$ 
\begin{equation*}
    \theta_{r} = \cos^{-1} \left(\frac{\cos \theta_{Y}}{\sqrt{1 + \mu_{s}^{2}}}\right) - \tan^{-1} \mu_{s} \ .
\end{equation*}
Note that $\theta_{r} \rightarrow 0^{\circ}$ as $\mu_{s} \rightarrow \infty$. 

Substituting $\alpha = 180^{\circ}$ into the general droplet profile equations of Section II, we get
\begin{equation} \label{X_p_prime}
    X_{p}'\left(Y\right) = \frac{f_{p} \left(Y\right)}{\sqrt{1 - \left\{f_{p} \left(Y\right)\right\}^{2}}}
\end{equation}
and
\begin{equation*}
    f_{p} \left(Y\right) = - \frac{1}{2} B \left(Y - \bar{Y}_{p}\right)^{2} + \bar{f}_{p},
\end{equation*}
\begin{equation*}
    \bar{Y}_{p} = \frac{1}{2} A^{*} - \frac{\sin \theta_{p}}{B}, \qquad \bar{f}_{p} = + \frac{1}{2} B \bar{Y}_{p}^{2} - \cos \theta_{p} \ .
\end{equation*}
We also have a relation
\begin{equation*}
    B Y_{p}^{2} - \left(A^{*} B - 2 \sin \theta_{p}\right) Y_{p} - \left(2 - 2 \cos \theta_{p}\right) = 0
\end{equation*}
whose solution is 
\begin{equation}
    Y_{p} = \bar{Y}_{p} + \sqrt{\bar{Y}_{p}^{2} + \frac{2}{B} \left(1 - \cos \theta_{p}\right)} \ . \label{eqn: Y_p}
\end{equation}
Note that $\bar{Y}_{p} \leq Y_{p}$, and $\bar{Y}_{p}$ is the inflection point for the pendent droplet profile. Unlike the case of sessile droplets, an inflection point can appear for pendent profiles when Bond numbers are not too small.

\begin{figure*}[!t]
  \includegraphics[width=0.9\textwidth]{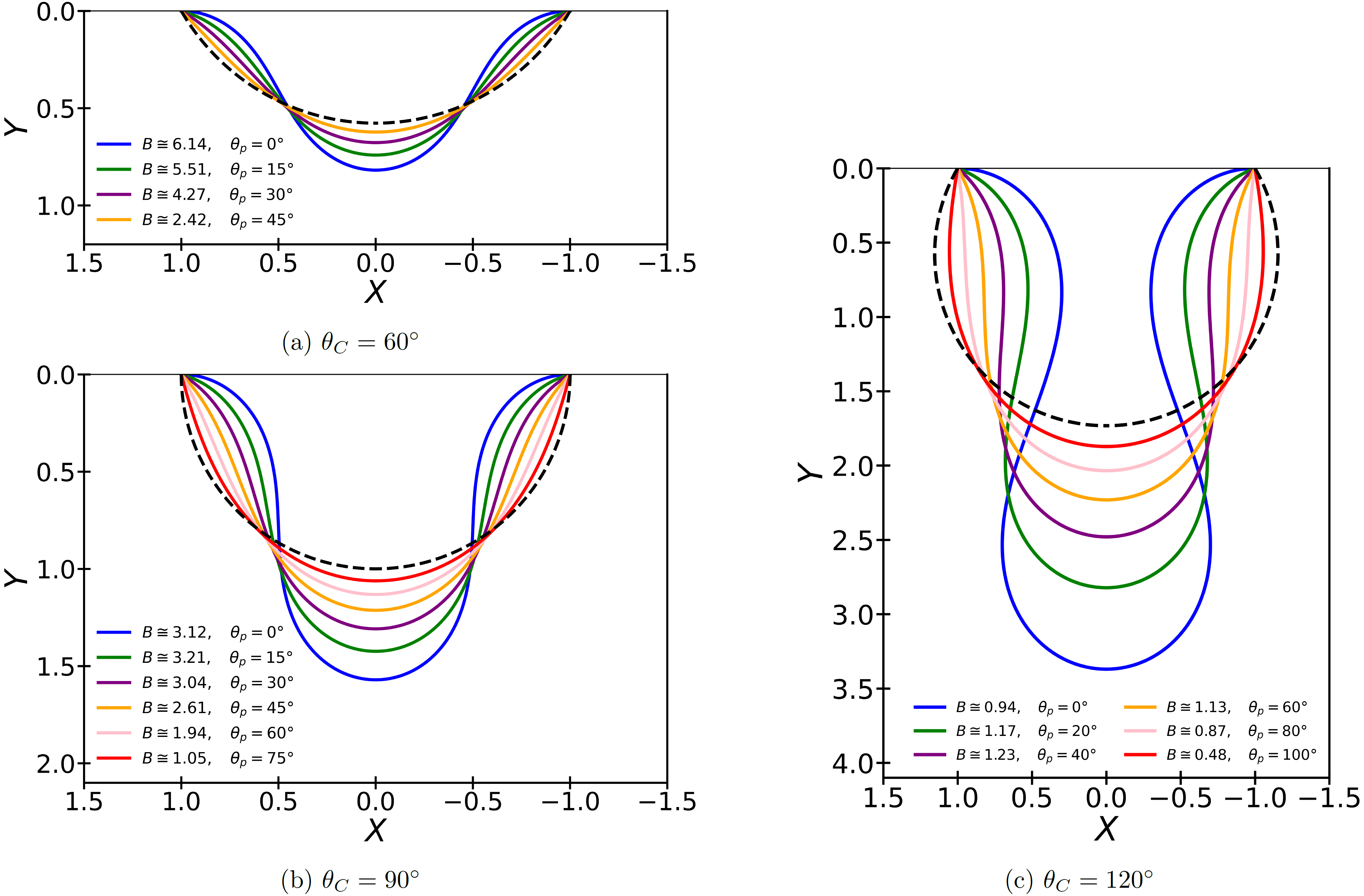}
  \caption{Shapes of pendent droplets under gravity.
           When $\theta_{C} \gtrsim 72.752^{\circ}$, there correspond two possible values of $\theta_{p}$ or $Y_{p}$ for a single Bond number $B$ (see Section VII).
           The black dashed lines are for $B = 0$ ($\theta_{p} = \theta_{C}$).}
  \label{Figure9}
\end{figure*}

The domain of $Y$ is either $\bar{Y}_{p} \leq 0 \leq Y \leq Y_{p}$ or $0 \leq Y \leq \bar{Y}_{p} \leq Y_{p}$ depending on the sign of $\bar{Y}_{p}$. $Y = \bar{Y}_{p}$ is the extreme point of $f_{p} \left(Y\right)$, and because it is convex upward, it represents the maximum point. When $\bar{Y}_{p} \geq 0$, the solution must be separately found for each domain of either $0 \leq Y \leq \bar{Y}_{p}$ or $\bar{Y}_{p} \leq Y \leq Y_{p}$. And if $\bar{Y}_{p} < 0$, the interval for $\bar{f}_{p}$ must be divided into $-1 < \bar{f}_{p} \leq 1$ and $1 < \bar{f}_{p}$ to find the solution. 

We now classify solutions according to the possible ranges of $\bar{Y}_{p}$ and $\bar{f}_{p}$ \cite{Dwight_1961}:

\noindent \underline{(1) $\bar{Y}_{p} \geq 0$ }

\noindent The droplet profile has the following solution
\begin{eqnarray*}
    X_{p} \left(Y\right) = 
    \begin{cases}
        \dfrac{1}{\sqrt{B}} \left[4 E \left(\dfrac{\pi}{2}, k\right) - 2 E \left(\phi, k\right)\right. \\
            \quad    \left.- 2 F \left(\dfrac{\pi}{2}, k\right) + F \left(\phi, k\right)\right] & \left(0 \leq Y \leq \bar{Y}_{p}\right) \\
        \\
        \dfrac{1}{\sqrt{B}} \left[2 E \left(\psi, k\right) - F \left(\psi, k\right)\right]     & \left(\bar{Y}_{p} \leq Y \leq Y_{p}\right) \ .
    \end{cases}
\end{eqnarray*}
The boundary condition at $X_{p} \left(0\right) = 1$ gives
\begin{align} \label{eqn: BC_pendent_1}
  \begin{split}
    \sqrt{B} = &4 E \left(\frac{\pi}{2}, k\right) - 2 E \left(\phi_{0}, k\right) \\
    &- 2 F \left(\frac{\pi}{2}, k\right) + F \left(\phi_{0}, k\right)
  \end{split}
\end{align}
where the arguments are related to Bond number $B$ and the contact angle $\theta_{p}$ as
\begin{align*}
    &\phi = \cos^{-1} \left(\frac{\sqrt{B}}{2k} \left(\bar{Y}_{p} - Y\right)\right), && \phi_{0} = \cos^{-1} \left(\frac{\sqrt{B}}{2k} \bar{Y}_{p}\right), \\
    &\psi = \cos^{-1} \left(\frac{\sqrt{B}}{2k} \left(Y - \bar{Y}_{p}\right)\right), && k = \sqrt{\frac{1 + \bar{f}_{p}}{2}} \ .
\end{align*}

\noindent \underline{(2) $\bar{Y}_{p} < 0$}

\noindent The profile solution is
\begin{eqnarray*}
    X_{p} \left(Y\right) = 
    \begin{cases}
        \dfrac{1}{\sqrt{B}} \left[2 E \left(\varphi, k\right) - F \left(\varphi, k\right)\right]                    \qquad \left(-1 < \bar{f}_{p} \leq 1\right)\\
        \\
        \dfrac{1}{\tilde{k} \sqrt{B}} \left[2 E \left(\chi, \tilde{k}\right)\right. \\ 
                         \qquad \quad \left.- \left(2 - \tilde{k}^{2}\right) F \left(\chi, \tilde{k}\right)\right]  \qquad \left(+1 < \bar{f}_{p}\right) \ .
    \end{cases}
\end{eqnarray*}
And the boundary condition gives
\begin{eqnarray} \label{eqn: BC_pendent_2}
    \sqrt{B} = 
    \begin{cases}
        2 E \left(\varphi_{0}, k\right) - F \left(\varphi_{0}, k\right)                                      \qquad \left(-1 < \bar{f}_{p} \leq 1\right) \\
        \\
        \dfrac{1}{\tilde{k}} \left[2 E \left(\chi_{0}, \tilde{k}\right)\right. \\
                      \quad \left.- \left(2 - \tilde{k}^{2}\right) F \left(\chi_{0}, \tilde{k}\right)\right] \qquad \left(+1 < \bar{f}_{p}\right)
    \end{cases}
\end{eqnarray}

\begin{figure*}[!t]
  \includegraphics[width=0.9\textwidth]{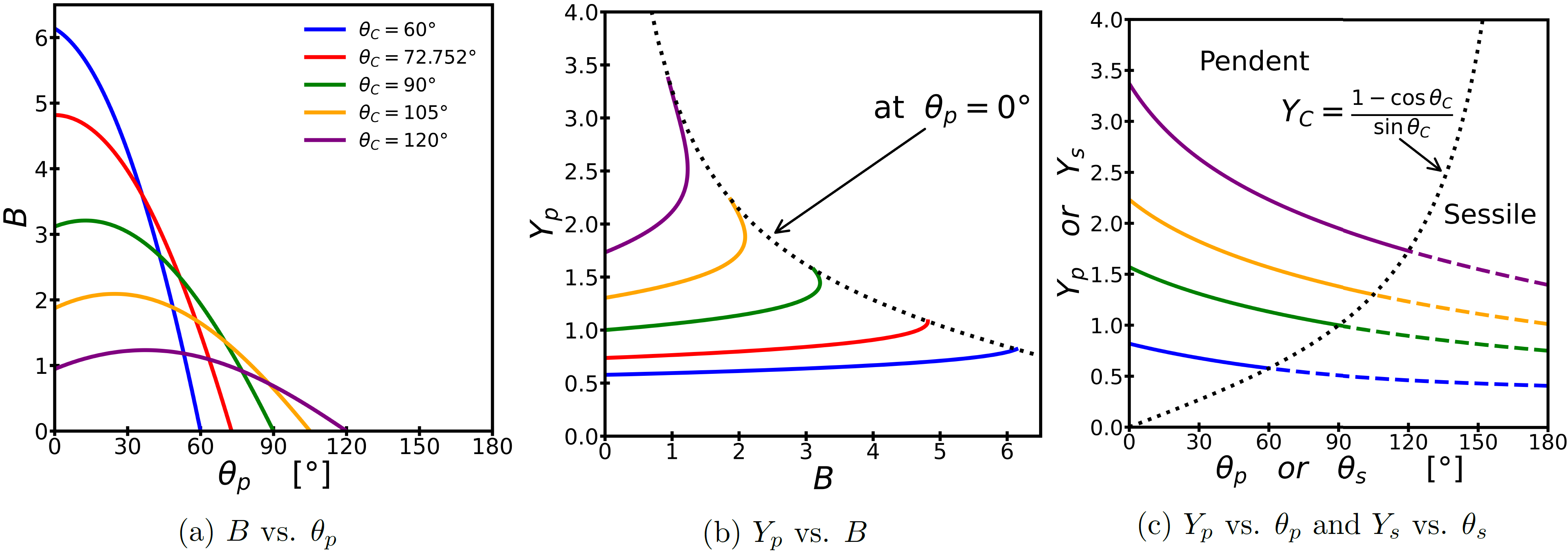}
  \caption{$\theta_{C} = 60^{\circ}$ (blue), $\theta_{C} = 90^{\circ}$ (green), $\theta_{C} = 105^{\circ}$ (orange), $\theta_{C} = 120^{\circ}$ (purple), and $\theta_{C} = 72.752^{\circ}$ (red). 
           (a) When $\theta_{C} \gtrsim 72.752^{\circ}$, there are two values of $\theta_{p}$ for a single $B$. 
           (b) Again for $\theta_{C} \gtrsim 72.752^{\circ}$, there correspond two values of $Y_{p}$ for a single $B$. 
           (c) The region above the curves $Y_{C}$ (dotted lines) is for the pendent droplets (solid lines), and the region below it is for the sessile droplets (dashed lines).}
  \label{Figure10}
\end{figure*}

\begin{figure*}[!t]
  \includegraphics[width=0.9\textwidth]{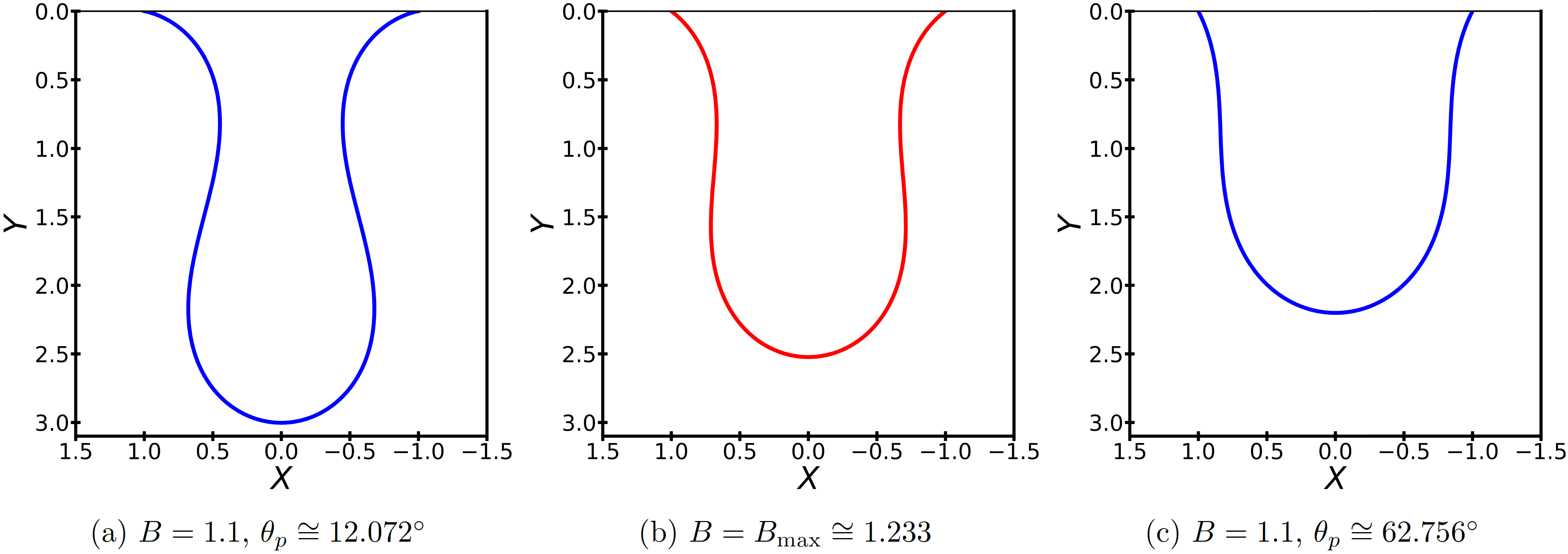}
  \caption{Two blue profiles are for $\theta_{C} = 120^{\circ}$ with the same Bond number ($B = 1.1$), and the red profile is for $\theta_{C} = 120^{\circ}$ with the maximum Bond number ($B = B_{\text{max}} \cong 1.233$, $\theta_{p} \cong 37^{\circ}$).
           (a) When $B$ decreases from $B = B_{\text{max}}$ to $B = 1.1$, $\theta_{p}$ decreases from $\theta_{p} \cong 37^{\circ}$ to $\theta_{p} \cong 12.072^{\circ}$.
           (b) When $B = B_{\text{max}} \cong 1.233$ and $\theta_{p} \cong 37^{\circ}$.
           (c) When $B$ decreases from $B = B_{\text{max}}$ to $B = 1.1$, $\theta_{p}$ increases from $\theta_{p} \cong 37^{\circ}$ to $\theta_{p} \cong 62.756^{\circ}$.}
  \label{Figure11}
\end{figure*}

\noindent where we have
\begin{align*}
    &\varphi = \cos^{-1} \left(\frac{\sqrt{B}}{2k} \left(Y - \bar{Y}_{p}\right)\right), \quad \varphi_{0} = \cos^{-1} \left(- \frac{\sqrt{B}}{2k} \bar{Y}_{p}\right), \\
    &\chi = \sin^{-1} \left(\frac{\sqrt{1 + f_{p} \left(Y\right)}}{\sqrt{2}}\right),    \quad \chi_{0} = \frac{\theta_{p}}{2}, \\
    &k = \sqrt{\frac{1 + \bar{f}_{p}}{2}}, \qquad \tilde{k} = \sqrt{\frac{2}{1 + \bar{f}_{p}}} \ .
\end{align*}

Using Eqs.(\ref{eqn: Y_p}) - (\ref{eqn: BC_pendent_2}), we can find the maximum height $Y_{p}$ and the pendent angle $\theta_{p}$ for given values of $\theta_{C}$ and Bond number $B$ (Fig.\ref{Figure10}). The shapes of pendent droplets for $\theta_{C} = 60^{\circ}$, $90^{\circ}$, and $120^{\circ}$ are shown in Fig.\ref{Figure9}.

\section{Bifurcation, or two distinct stable pendent profiles}
For a sessile droplet with a fixed Bond number the profile solution is unique, implying we have unique values of the contact angle $\theta_{s}$ and the maximum height $Y_{s}$. This need not be true for a pendent droplet. With the help of numerical analysis we have found that there could be two stable profiles for a given Bond number. Two sets of contact angle $\theta_{p}$ and maximum heights $Y_{p}$ are possible for a fixed Bond number. The range of $\theta_{C}$ in which two solutions appear for a single $B$ is $72.752^{\circ} \lesssim \theta_{C} \leq 180^{\circ}$, where the lower bound of $\theta_{C}$ comes from the condition
\begin{equation*}
    \left.\frac{\partial B}{\partial \theta_{p}}\right|_{\theta_{p} = 0^{\circ}} = 0 \qquad \text{at} \quad \theta_{C} \cong 72.752^{\circ} \ .
\end{equation*}

We show graphs of solutions for various combinations of parameters in Fig.\ref{Figure10}. We also plotted two solutions for a given Bond number ($B=1.1$, $\theta_{C} = 120^{\circ}$) in Fig.\ref{Figure11}. In a very highly constrained experimental setting we might be able to massage a pendent droplet smoothly from one stable profile into another. Though it would be very hard to implement in real settings, it seems to be very interesting to consider the possibility to observe it.

When $\theta_{C} \lesssim 72.752^{\circ}$, if we increase the Bond number the contact angle converges to zero. This implies the receding angle has been already crossed and the contact points should move to achieve a new equilibrium profile. However, this is not the case for $\theta_{C} \gtrsim 72.752^{\circ}$. As we increase $B$ beyond its maximum value $B_{\text{max}}$ which is attained at a non-zero value of $\theta_{C}$ which could be larger than the receding angle ($\theta_{p} > \theta_{r}$), the contact points do not move, but we do not have a stable static solution since $B > B_{\text{max}}$. We guess that it corresponds to a droplet 'dynamically' pinching off a part of its mass and settling to a new stable configuration with the remaining part of the droplet. In the next section we will discuss a different kind of pinch-offs in which the size of a neck of a pendent droplet shrinks below the capillary length.

\section{Pinch-off for pendent droplets}
In the case of pendent droplets pinch-offs can occur even when $B < B_{\text{max}}$. In the below we will use the suffix 'p-o' to denote all the parameters just at the moment of pinch-off. In some recent studies it was pointed out that there could be pinch-offs for two-dimensional pendent droplets \cite{Liu_2014, Grinfeld_2024}. Their setting is different from ours in that the contact angle is a free parameter since they did not take account of contact angle hysteresis. If we take contact angle hysteresis into account, the pendent angle $\theta_{p}$ (i.e. the contact angle for pendent droplets) is not a free parameter because it is intricately intertwined with other parameters such as $\theta_{C}$, $B$, and $Y_{p}$, etc.

\begin{figure*}[!t]
  \includegraphics[width=0.9\textwidth]{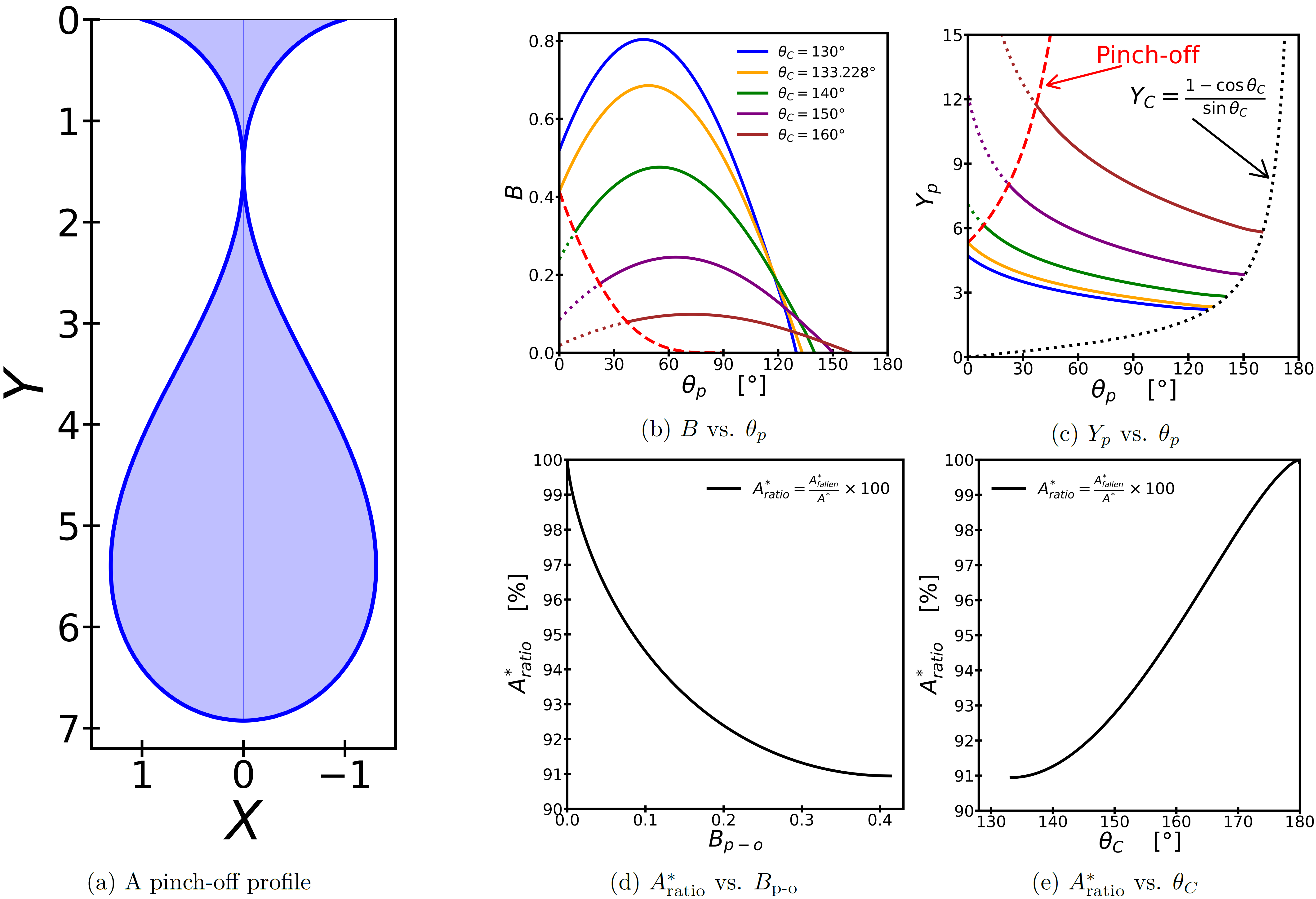}
  \caption{(a) The profile of a pendent droplet at the moment of pinch-off when $\theta_{p} = \theta_{\text{p-o}} = 15^{\circ}$.
          (b) and (c): The dotted graphs in color represent the regions forbidden because of pinch-offs. 
          $\theta_{C} = 130^{\circ}$ (blue), $\theta_{C} = 133.228^{\circ}$ (orange), $\theta_{C} = 140^{\circ}$ (green), $\theta_{C} = 150^{\circ}$ (purple), and $\theta_{C} = 160^{\circ}$ (brown). The red dashed line is the critical boundary beyond which there is no stable configurations. The black dotted line is for the case of no gravity ($B = 0$).
          (d) and (e): The ratio of the fallen-out to the total area.}
  \label{Figure12}
\end{figure*}

If the pendent droplet has an inflection point (i.e. $\bar{Y}_{p} \geq 0$), there exist both a neck and a bulge in the profile (see Fig.\ref{Figure9} (c) or Fig.\ref{Figure12} (a)). The positions of a neck or a bulge $\left(X_{p, \pm}, Y_{p, \pm}\right)$ are defined by the relation $X_{p}' \left(Y_{p, \pm}\right) = 0$ where $X_{p, \pm} \equiv X_{p} \left(Y_{p, \pm}\right)$. $(+)$ and $(-)$ signs are for a bulge and a neck, respectively. Then we have
\begin{equation*}
    Y_{p, \pm} = \bar{Y}_{p} \pm \sqrt{\frac{2}{B} \bar{f}_{p}} \ .
\end{equation*}

A pinch-off occurs when the position of a neck becomes zero: $X_{p, -} = 0$. We then get the following relation
\begin{equation*}
    \begin{split}
        0 = &4 E \left(\dfrac{\pi}{2}, k_{\text{p-o}}\right) - 2 E \left(\phi_{-}, k_{\text{p-o}}\right) \\
            &- 2 F \left(\dfrac{\pi}{2}, k_{\text{p-o}}\right) + F \left(\phi_{-}, k_{\text{p-o}}\right)
    \end{split}
\end{equation*}
where
\begin{equation*}
    \phi_{-} = \sin^{-1} \left(\frac{1}{\sqrt{2} k_{\text{p-o}}}\right) \ .
\end{equation*}
Solving this equation numerically, we get the value $k_{\text{p-o}} \cong 0.8551$. We find that it agrees well with Liu et al.'s result \cite{Liu_2014}.

Next, we can find the Bond number $B_{\text{p-o}}$ and the pendent angle $\theta_{\text{p-o}}$ at the pinch-off.
Combining $\bar{f}_{p} = B \bar{Y}_{p}^{2} / 2 - \cos \theta_{p}$ and $k = \sqrt{\left(1 + \bar{f}_{p}\right) / 2}$, we get a relation 
\begin{align} \label{eqn: condition_1}
    \begin{split}
        0 = &\left(A^{*} B_{\text{p-o}} - 2 \sin \theta_{\text{p-o}}\right)^{2}  \\
            &- 8 B_{\text{p-o}} \left(2 k_{\text{p-o}}^{2} - 1 + \cos \theta_{\text{p-o}}\right) \ .
    \end{split}
\end{align} 
And from the boundary condition Eq.(\ref{eqn: BC_pendent_1}) we obtain another relation
\begin{equation*}
    \begin{split}
        \sqrt{B_{\text{p-o}}} = &4 E \left(\frac{\pi}{2}, k_{\text{p-o}}\right) - 2 E \left(\phi_{\text{p-o}}, k_{\text{p-o}}\right) \\
                                &- 2 F \left(\frac{\pi}{2}, k_{\text{p-o}}\right) + F \left(\phi_{\text{p-o}}, k_{\text{p-o}}\right)
    \end{split}
\end{equation*}
where
\begin{equation*}
    \phi_{\text{p-o}} = \sin^{-1} \left(\frac{\sqrt{1 - \cos \theta_{\text{p-o}}}}{\sqrt{2} k_{\text{p-o}}}\right) \ .
\end{equation*}
We can find $B_{\text{p-o}}$ and $\theta_{\text{p-o}}$ satisfying these two relations for given $A^{*}$ (or equivalently $\theta_{C}$).

Note that $\theta_{\text{p-o}} \rightarrow 90^{\circ}$ as $\theta_{C} \rightarrow 180^{\circ}$ in Eq.(\ref{eqn: condition_1}). Since there are at most two extremum points for a two-dimensional pendent droplet (Eq.(\ref{X_p_prime})), the angle of pinch-off $\theta_{\text{p-o}}$ should be less than $90^\circ$.

The range of $\theta_{C}$ in which a pinch-off occurs is $133.228^{\circ} \lesssim \theta_{C} < 180^{\circ}$, where the lower bound comes from the conditions when $\theta_{\text{p-o}} = 0^{\circ}$. Thus, there appears the range of $B$, $\theta_{p}$, and $Y_{p}$ which is forbidden due to pinch-offs (see Fig.\ref{Figure12} (b) and (c)). In this sense $\theta_{\text{p-o}}$ can be interpreted as the minimum value of the pendent angle for a pinch-off not to occur.

If we solve Eq.(\ref{eqn: condition_1}) for $A^{*}$, we obtain
\begin{equation*}
    A^{*} = \frac{2 \sin \theta_{\text{p-o}}}{B_{\text{p-o}}} + \frac{2 \sqrt{2}}{\sqrt{B_{\text{p-o}}}} \sqrt{2 k_{\text{p-o}}^{2} - 1 + \cos \theta_{\text{p-o}}} \ .
\end{equation*}
The total area $A^*$ is the sum of the fallen-out area $A_{\text{fallen}}^{*}$ and the remnant area $A_{\text{remnant}}^{*}$
\begin{equation*}
    A^{*} \equiv A_{\text{remnant}}^{*} + A_{\text{fallen}}^{*} \ .
\end{equation*}
By integrating the profile function for the dangling droplet below the neck we can find the fallen area $A_{\text{fallen}}^{*}$. We obtain a surprisingly simple relation between $A_{\text{fallen}}^{*}$ and $B_{\text{p-o}}$:
\begin{equation*}
    A_{\text{fallen}}^{*} = \frac{2}{B_{\text{p-o}}} \ .
\end{equation*}
In the limit of $B_{\text{p-o}} \rightarrow 0$, it is obvious that $A^{*} \rightarrow \infty$ (or equivalently, as $\theta_{C} \rightarrow 180^{\circ}$). Then it requires $\theta_{\text{p-o}} \rightarrow 90^{\circ}$ as we argued in the above. Hence, it means $A_{\text{fallen}}^{*} \rightarrow A^{*}$. Intuitively this is correct since the larger the size of the pendent droplet, the more likely it is to be fallen even by a small gravitational pull with a negligible remnant. From the numerical works we found that the remnant area after the pinch-off does not exceed approximately $9\%$ of the area before the pinch-off (see Fig.\ref{Figure12} (d) and (e)).

\section{Conclusion}
In this work we have examined the static friction between liquid droplet and solid substrate to understand and analyze the problem of contact angle hysteresis. We determined the profiles of 2-dimensional droplets in various and general settings very thoroughly and systematically. We have derived the modified Young's equation and the Young-Laplace's equations using energy minimization under two constraint conditions of mass conservation and fixed contact points. The constraint forces are identified as the total normal force (or the total suction for some cases of pendent droplets) and the static friction at contact points. The contact angle hysteresis and the pinning force can be explained in the context of static friction. For instance, the Furmidge relation can be understood as a very special case within a more general framework.

By putting the droplet on an inclined plane of which angle is a free adjustable parameter, we could have understood the shift of profiles under changes of inclination angle and Bond number. While the previous works on droplets have been done separately by bits and pieces, we have unified the discussion of all kinds of 2-dimensional droplets, sessile or pendent, in a single framework.

We also found that there could be two distinct stable profile solutions for a given Bond number under a certain condition, {\it bifurcation}. This observation has been made for the first time, and its implications have been discussed briefly. 

For the case of pendent droplets we have discussed in detail how a pinch-off can occur for given parameters. Among our findings there is a very interesting relation between the fallen-out area and the Bond number at the moment of pinch-off. We are not aware of any previous works that pointed out this.

Though this work is concerning only 2-dimensional droplets, which is not easy to implement experimentally, this limitation can be remedied by considering the systems like capillary bridges which are innately 2-dimensional and are experimentally more feasible. This topic has been discussed in our subsequent work \cite{Yang_arXiv_Bridge_2024}. Various aspects of contact angle hysteresis that we found in this work have been considered there with the experimental implementations in mind.

\begin{acknowledgments}
One of the authors (J.-I. Y) would like to thank Profs. Bo Soo Kang and Young-Dae Jung for their support and encouragement throughout the course of his graduate years.
\end{acknowledgments}

\nocite{*}

\bibliography{Droplet_refs.bib}

\end{document}